\documentclass[letterpaper, 10 pt, conference]{ieeeconf}  

\IEEEoverridecommandlockouts                              

\title{\LARGE \bf
Trust in robot-mediated health information*
}

\author{David Cameron$^{1}$, Marina Sarda Gou,$^{2}$ and Laura Sbaffi$^{1}$
\thanks{*This work was funded by The University of Sheffield, Information School's researcher development fund}
\thanks{$^{1}$Information School, The University of Sheffield, Sheffield, UK
        {\tt\small [d.s.cameron] [l.sbaffi]@sheffield.ac.uk}}%
\thanks{$^{2}$Department of Psychology, The University of Sheffield, Sheffield, UK
        {\tt\small msardagou1@sheffield.ac.uk}}%
}

\begin{document}

\maketitle
\thispagestyle{empty}
\pagestyle{empty}

\begin{abstract}

This paper outlines a social robot platform for providing health information. In comparison with previous findings for accessing information online, the use of a social robot may affect which factors users consider important when evaluating the trustworthiness of health information provided.

\end{abstract}

\section{INTRODUCTION}



Ahead of the use of social robotics healthcare and health information contexts, it is important to identify what prospective health information users would want from such services. Technology is advancing rapidly and health information is increasingly mediated by computers and accessed online \cite{rowley2015students}; there may be valuable insights from the current use of online health information to be considered in the further introduction of social robotics in health care.

Although online websites and social robots have potential to deliver accurate health information, the interaction experience for each of these may be quite different. In comparison with virtual agents, robots are rated as having more sociability, responsiveness, and trustworthiness \cite{powers2007comparing}; a dynamic and responsive agent could also be seen as a social entity and treated as if it was a real person \cite{reeves1996media}. There is already some evidence of these differences in the contexts of educational information \cite{hsu2007investigation}. Thus, people seeking health information might not have the same reactions interacting with static information online in comparison to a social robot; this may affect how they evaluate information provided.

In the present article, we specifically consider the issue of trust. Trust in robots is still an emerging area in the field of Human-Robot Interaction (HRI) \cite{Schaefer2016trust} but nonetheless is increasingly recognised as influencing the course of interaction, such as people's willingness to follow advice \cite{freedy2007measurement}. Trust in HRI has a contextual basis \cite{cameron2015framing}, so the nature of the interaction - in this case receiving health information - may specifically frame people's trust towards a social robot in ways that resemble other health information contexts (e.g., from a pharmacist or from health websites).

\subsection{Trusting Health Information}

Understanding how people come to trust the information and advice they find online has been an important issue since the widespread adoption of digital technologies \cite{sillence2007patients}. Despite the introduction of standards, concerns over information quality, accuracy and credibility 
are still echoed by researchers examining the provision of health information material across a range of conditions \cite{daraz2018readability}. Recent research has demonstrated that there are many factors to consider when addressing trust formation in health information, all of which are interdependent and interconnected \cite{rowley2015students}. 

\subsection{Research Aim}

This exploratory work examines if the use of a social robot platform influences which factors users consider important in evaluating the trustworthiness of health information offered. Outcomes in this study are compared against previous findings for users searching for health information online \cite{rowley2015students}.

\section{Method}

A brief, staged, video scenario was shown to represent a potential use of robots in healthcare. In the video, Pepper first introduced itself as a source of health information. A person then asks Pepper for advice on how to treat a minor injury (knee pain) and Pepper delivers advice through synthesised speech. All advice came from the relevant NHS website.

Trust questions were drawn from a scale used to establish trust in online health information \cite{rowley2015students}. Questions were adapted to reflect the use of a social robot as an interface for health information (e.g., `The extent to which the \textit{article} gives me information that I can use' became `The extent to which the \textit{robot} gives me information that I can use'). 

As with the original study \cite{rowley2015students}, questions were presented as a bank of 5-point Likert-style statements to investigate participants' reports of the relative importance of each statement in determining a health-information robot's trustworthiness.

Participants were recruited through the University of Sheffield volunteers list. 97 participants (34 identified as male, 63 as female; Mean age = 32.78) passed the attention screening test (accurately describing the health issue presented in the video) and completed the questionnaire. Participants were invited to take part in a prize draw as recompense for their time.

\section{Results}

Responses to the questions were collated across the 8 main factors previously identified by students for trusting online health information \cite{rowley2015students}. Mean and SD scores for these factors are reported in Table 1 and factors are ranked by their mean importance.

\begin{table}[h]
\centering
\caption{factors ranked by reported importance}
\begin{tabular}{llrr}
\hline
factor         & factor Definition                                                                                       & Mean & SD   \\ \hline
Style          & \begin{tabular}[c]{@{}l@{}}The way information \\ is presented and written\end{tabular}          & 4.30 & \phantom{0}.72 \\
Credibility    & \begin{tabular}[c]{@{}l@{}}Information is believable\\ and impartial.\end{tabular}               & 4.29 & \phantom{0}.73 \\
Ease of use    & \begin{tabular}[c]{@{}l@{}}Ease of accessing and \\ using the robot\end{tabular}                 & 4.20 & \phantom{0}.72 \\
Content        & \begin{tabular}[c]{@{}l@{}}Information's characteristics:\\  reliability, accuracy, currency\end{tabular} & 4.15 & \phantom{0}.82 \\
Usefulness     & \begin{tabular}[c]{@{}l@{}}Extent the user can make \\ use of the information\end{tabular}       & 4.09 & \phantom{0}.83 \\
Verification   & \begin{tabular}[c]{@{}l@{}}Apparent expertise and \\ consistency\end{tabular}                    & 3.86 & \phantom{0}.94 \\
Recommendation & \begin{tabular}[c]{@{}l@{}}Recommendations from \\ known persons\end{tabular}                    & 3.26 & 1.11 \\
Brand          & \begin{tabular}[c]{@{}l@{}}Robot brand indicators\\ and reputation\end{tabular}                  & 3.04 & 1.03 \\ \hline
\end{tabular}
\end{table}

\subsection{Comparison with historical data}

The ranked importance of factors for the current results differs substantially from historical data for the factor rankings for online health information. Table 2 highlights the relative importance in terms of rank for each factor used in determining the trustworthiness of health information online or delivered by a robot platform.

\begin{table}[h]
\centering
\caption{factor rankings compared with historical data}
\begin{tabular}{llc}
\hline
Historical Ranking {[}1{]} & Current Ranking & Change    \\ \hline
Credibility           & Style           & $\uparrow   2$ \\
Content               & Credibility     & $\downarrow 1$ \\
Style                 & Ease of use     & $\uparrow   4$ \\
Usefulness            & Content         & $\downarrow 2$ \\
Brand                 & Usefulness      & $\downarrow 1$ \\
Verification          & Verification    & --             \\
Ease of use           & Recommendation  & $\uparrow   1$ \\
Recommendation        & Brand           & $\downarrow 3$ \\ \hline
\end{tabular}
\end{table}

\section{Discussion}

The results indicate that the use of a social robot as a platform for health information may influence people's views on what to consider when it comes to assessing the trustworthiness of the information. Specifically, in comparison to historical data for people assessing online delivery of health information \cite{rowley2015students}, robot-mediated information may result in people prioritising different aspects of such information.

In the current work, style is recognised as the most important aspect of assessing trustworthiness. Given the unfamiliarity most people would have in interacting with a social robot compared to their experience of using sites online, this may reflect clarity as an important issue in (at least initial) interaction \cite{rossi2018getting}. Similarly, ease of use is ranked as higher importance in the current study than in the previous. Again, when facing new health technology, ease of use is a significant predictor in its acceptance \cite{pai2011applying}; the novelty of social robots in this context could potentially inflate the importance of ease of use. 

In contrast, a robot's branding is ranked as least important for determining trustworthiness. Without established reputations to associate with different robots or manufacturer brands, a robot's brand may provide very little information to users; this may present an opportunity for health information providers to attach their own brands to such devices.

\subsection{Limitations}
The are distinct differences in focus and methodology between this and the prior study \cite{rowley2015students}; caution is advised when interpreting these early results. The previous work comprised participants considering their own active search for health information on existing technology, whereas this study asks participants to observe a novel interaction and imagine. Nonetheless, the results point to potentially viable areas for further research.

\subsection{Future directions}

Of note, credibility is ranked high across both online and robot-mediated health information; trust in these health contexts may have a significant social element - the information is perceived to be provided with the user's interest and well-being in mind - as seen in face to face healthcare \cite{croker2013factors}. Given users may perceive robots as social agents, and the specific social contexts in which health information may be accessed, it may be productive to further explore this factor in terms of trust in robots as a social construct.


\addtolength{\textheight}{-12cm}   

\bibliographystyle{IEEEtran}
\bibliography{name}


\end{document}